\documentclass[10pt,conference,compsocconf]{IEEEtran}
\usepackage{times}
\usepackage{amsmath, amsfonts}
\usepackage{caption}
\usepackage{algorithm}
\usepackage[noend]{algpseudocode}
\usepackage{graphicx}
\ifCLASSINFOpdf
\else
\fi
\makeatletter
\def\BState{\State\hskip-\ALG@thistlm}
\makeatother

\begin{document}
\pagenumbering{gobble}
%
\title{\textbf{\Large Implementing Suffix Array Algorithm Using Apache Big Table Data Implementation}\\[0.2ex]}

\author{
\IEEEauthorblockN{~\\[-0.4ex]\large Piero Giacomelli\\[0.3ex]\normalsize}
\IEEEauthorblockA{Fidia Farmaceutici S.p.A\\
36075 Abano Terme, Padova Italy\\
Email: {\tt pgiacomellip@fidiapharma.it}
}
}

%


\maketitle

\begin{abstract}
In this paper we will describe a new approach on the well-known suffix-array algorithm using Big Table Data Technology. We will demonstrate how it is possible to refactor a well-known algorithm coupled by taking advantage of an high-performance distributed datastore, to illustrate the advantages of using datastore cloud related technology for storing  large text sequences and retrieving them. A case study using DNA strings, considered one of the most difficult pattern matching problem, will be described and evaluated to demonstrate the potentiality of this implementation. Further discussion on performances and other big data related issues will be described as well as new possible lines of research in big data technology for precise medical applications.
\end{abstract}


\begin{IEEEkeywords}
Suffix Array; Big Table Data; Apache Accumulo%
\end{IEEEkeywords}

Implementing fast string pattern matching algorithm to perform fast data retrieval in large text sequence is something crucial for database development and text processing software, and it's becoming more and more important as the availability of text data production increases at current rate (see for example \cite{mislove2008growth}). This paper will illustrate the possibility offered by implementing suffix array algorithm using Big Table data storage. The paper is organized as follows: in section I we will introduce the general problem. Section II will describe and state the problem definition while section III will review the suffix array algorithm. Section VI will describe in deep the implementation of the suffix array algorithm relying on the Apache Accumulo datastore. In section V we will describe our performance analysis and at the end we will detail our conclusion and future implementations.
\section{Introduction}
Even if it is difficult to evaluate the growth of DNA sequence available mainly due to the fact that DNA sequence are stored in non-public repositories, the birth of private companies that collect DNA sequence form the saliva of patient to give them report of possible predispositions to some disease like 23andme \cite{23andme}, and public genomic repositories give boost to the so called "personalized medicine".
So the fast DNA data are available the fast the need of efficient pattern matching algorithm over cloud based repositories is becoming an urgent problem to face with. Traditional Remote Database Management system (RDMS) has demonstrate their limit in handling DNA sequence and they lack in performance when using (Structured Query Language) SQL to perform CRUD(Create, Read, Update and Delete) operation over on-structured DNA string \cite{}. In general, having one large text where it is needed to perform pattern matching is one of the oldest algorithm that have been implemented. Only to give and hint to the reader, actually the growth rate of the Wikipedia dataset is between 1 and 1.4 GB a week \cite{wikipedia_grow}. Considering that, a single human complete DNA sequence is half of this size one can have the idea of how big DNA dataset could be. Despite the fact that pattern matching algorithm is one of the oldest branch in computer science literature, the problem of storing  DNA sequences and performing pattern matching on them, using also parallelism is still object of research\cite{faro2013exact}. The advent of the cloud paradigm and the use of large distributed server farms to store information has given the opportunity to refactor and recode these algorithm in a new way. In this paper, we will describe a new way to store DNA sequence using Big Data table technology and using this as dataset we will recode the suffix array algorithm to demonstrate the possible advantages of this approach compared to the traditional one.

\section{Problem definition}
Before moving on we will briefly review the main algorithm we are going to use in the paper against a Big Table dataset. From a general point of view string pattern matching could be, formally speaking, stated as follows:
 let $A=\{a_1..a_n\}$ be and alphabet i.e. a finite number of symbols (letters). We define also the lexicographic order over $A$ by defining that $a_1\leq a_2 \cdots \leq a_n$ and so on. A word $W$ is defined as an element of the set $A*=\{a_1a_2.. | a_i \in A,  i \in \mathbb{N}\}$. The length of a word $w$ is the number of letters the word is formed of and a word $w$ is finite if $i$ is finite. So giving a word $w_N$ of length $N$ and a word $W_k$ of length $k$, Where $k << N$ the problem of finding a pattern matching can be stated as follows: find it there is an index $j_0 \in \mathbb{N}$ such that
\begin{equation*}
\forall i=j_0..j_0+k, a_i=w_i.
\end{equation*}
This is the simplest problem that can be stated as pattern matching but also find all the occurrences of a given word into a larger text is also important. The standard reference is the classic work of Knuth \cite{knuth1977fast}, for a more modern approach in DNA sequence we suggest \cite{faro2013exact}. 
Even if, various methods for solving this problems have been proposed, still some research needs to be done to see how it is possible to efficiently implement the algorithms for solving the pattern matching problem, when the original string is not stored in one single storage  but it is distributed over different nodes that so need to handle unavailability and failure in retrieving. Before the booming of the social networks searching string in long text was linked to biological DNA sequence. From a string pattern matching point of view, the Homo Sapiens Sapiens DNA first sequenced with the Human Genome Project \cite{collins2003human},  can be considered as a string composed of the basic four basic letter nuclotides. Each nucleotide is composed of a nitrogen-containing nucleobase—either cytosine (C), guanine (G), adenine (A), or thymine (T). So the whole Human DNA can be thought as a string based on four letter whose length is approximately \cite{venter2001sequence} $3.2$ billion base pairs where each base pair is the letter identifying one of the nucleotides. So a portion of the string looks something like 
\newline
\newline
AGCCCCTCAGGAGTCCGGCCACAT
\newline
\newline
If we use two bits for identify each letter, one can use $00$, $01$, $10$, and $11$ for $T$, $G$, $C$ and $A$, we have that the whole Human DNA is $3.2 \cdot 2=6.4 $ billions bit that is something like $800$ megabytes. This number is realted only for one human being without considering any loss on the string due to missing information. So right now the whole DNA could be easily stored into a common USB flash stick. However searching a pattern on this big string is an high computational task, not considering the fact that if we need to perform a string pattern matching over different DNA sequences is is computationally unhandleable using a brute force approach. 
\section{Suffix-array algorithm}
We are ready now to review how the suffix-array algorithm works. As we have seen in the problem definition, the easier way of searching a a pattern string within a bigger string of this size could be using a brute force attack. The simplest algorithm is illustrated using the pseudo-code below

\begin{algorithm}
\caption{Substring search algorithm}\label{alg:sssa}
\begin{algorithmic}[1]
\Procedure{BruteForceSearch}{$pattern, subject $}
\State $n \gets len(subject)$
\State $m \gets len(pattern)$
\For{$i = 0; i < n-m; i++ $}
\State $j \gets 0$
\While{$j < m$ and  $subject_{i+j} = pattern_j$}  
	\State $j \gets j+1$
\EndWhile
\If{$j=m$} 
	\Return $i$ 
\EndIf
\EndFor
	\Return $-1$ 
\EndProcedure
\end{algorithmic}
\end{algorithm}

So, as one may see, the running time of this algorithm depends upon the position of the searched string within the whole string so, giving a large string of size $N$ and a pattern to search containing $k$ letters, the total number of comparison that should be done are $k(N-k+1)$. In the worst case, i.e. the pattern could not be find in the larger string is $O(Nk)$ we need to perform at least $N-k$ comparisons on the base string . During the years an intense literature have been developed to find algorithms for string pattern matching that are more time and space efficient \cite{singla2012string} \cite{bezdek2013pattern}, \cite{gawrychowski2013optimal}.
Suffix array \cite{karkkainen2003simple} is a well-known algorithm on the field of pattern-matching over big strings. The algorithm itself, is relatively old and have been extensively studied \cite{manber1993suffix}. Its counterpart algorithm using suffix-tree \cite{mccreight1976space} is known to be the fastest way to search occurrence of a given string over big strings. From an high point of view the two algorithms work the same, being that they do not perfom the comparisons over the original string, but instead, from the original string they construct a new data structure where it is easier to search a string or counting its occurrences.
In particular the suffix-array algorithm is organized in two stages, in the first part we consider the original string and we create all the suffices from the original string. Then we constructed an ordered set using the lexicographic order of the alphabet into the suffices. Now that we have completed this stage we have and ordered set of suffices where we can now do our search for pattern matching, using more efficient algorithm in particular binary search tree \cite{bentley1975multidimensional}. 
To illustrate the algorithm we consider the classical example of the string "MISSISSIPPI". 
The suffices are 
$\{$ MISSISSIPPI, ISSISSIPPI, SSISSIPPI, SISSIPPI, ISSIPPI, SSIPPI, SIPPI, IPPI
, PPI , PI, I $\}$. If we order them into a lexicographic set we have the following order array
\newline
I  \newline
,IPPI  \newline
,ISSIPPI  \newline
,ISSISSIPPI  \newline
,MISSISSIPPI  \newline
,PI  \newline
,PPI  \newline
,SIPPI  \newline
,SISSIPPI  \newline
,SSISSIPPI  \newline
\newline
	Now using this ordered set it is possible to find the best match using a divide and conquer approach \cite{bentley1997fast}.
Using pseudo-code we can code the suffix array algorithm the following way:
\begin{algorithm}
\caption{suffix array pattern matching}\label{alg:sssa}
\begin{algorithmic}[1]

\Procedure{Build suffix array}{$text$}
\State $n \gets len(text)$
\State $suffices \gets \emptyset$
\For{$i = 0; i < n; i++ $}
\State $suffices_i \gets text_{i,n}$
\EndFor
\EndProcedure
\Procedure{Search suffix array}{$pattern, suffices$}
	\If{$match$} 
		\Return $suffices[i]$         
	\EndIf
	\Return $\emptyset$     
\EndProcedure
\end{algorithmic}
\end{algorithm}

One good implementation of the the suffix array algorithm 
\cite{jsuffixarrays} is freely available to the public audience at the source code repository GitHub \cite{github}. This simpler approach could also be tuned to perform less comparisons, moving from the fact that giving an alphabet $A=\{a_1..a_n\}$ of cardinality $n$ the starting letter of the suffices are exactly $n$ and so we can faster our search by using in a first stage only the comparison of the first letter of the substring against the suffix array and then once we have found a match starting with the same procedure to see if a match exists. This search can be done using the divide and conquer approach where the number of possibles comparison decrease with an exponential factor of $2$. In our example if we need to find the match to the string $PI$ we perform a maximum of $1$ comparison to find the only suffix that start with the letter $P$ and then 1 additional comparison to find the that the world PI is contained in the world MISSISSIPI, compared to the 11 that are needed on the brute-force approach.
Now that we introduced the original algorithm we move on by describing our implementation based on Big Table technology.
\section{Accumulo Implementation of Suffix array}
Apache Accumulo \cite{halldorsson2013apache} is a highly scalable, distributed, open source data store modeled after Google’s Bigtable design \cite{chang2008bigtable} . 

	Accumulo is built to store up to trillions of data elements and keeps them organized so that users can perform fast lookups. Accumulo supports flexible data schemas and scales horizontally across thousands of machines. Applications built on Accumulo are capable of serving a large number of users and can process many requests per second, making Accumulo an ideal choice for terabyte- to petabyte-scale projects.   
    
	Accumulo employs this distributed approach by partitioning data across multiple servers and keeping track of which server has which partition. In some cases these data partitions are called shards, as in pieces of something that has been shattered. 
    
    In Accumulo' s case, data is stored in tables, and tables are partitioned into tablets. Each server hosts a number of tablets. These servers are called tablet servers.
	The idea behind our implementation is to use Accumulo's tablets to store portion of a DNA string. Once we finish the pre-processor phase we will use the Accumulo algorithm for solving our pattern matching problem.    
	In our approach we code a Java Accumulo client class that accept as input the complete source of the first chromosome of the Homo Sapiens Sapiens (freely available at \cite{faro2013exact}) and we transform every suffix of it into a row of an Accumulo table by assign as unique ROWID the starting position of the suffix to be stored. So the position where the suffix started is the rowid of the corresponding Accumulo tablets.
    
    Being that the first suffices are basically the whole string minus some characters, and being that we are interested in finding short string inside bigger strings we limit the suffices to the first 1000 characters. So every our final Accumulo table is shown in TABLE I.
    
\begin{table}[!htbp] \centering 
  \caption{DNA row of Accumulo table} 
  \label{} 
\begin{tabular}{@{\extracolsep{5pt}} cccccc} 
\\[-1.8ex]\hline 
\hline \\[-1.8ex] 
 ROWID & TEXT \\ 
\hline \\[-1.8ex] 
1 & CTACTGACTGCTGTGTGGGTTATCTACTA \\ 
2 & CTGACTCTATATAGTTACGTACCCGGTGA \\ 
3 & TAGTGCAGCTTCTATCTAAAGATCGCCGT \\ 
\hline \\[-1.8ex] 
\end{tabular} 
\end{table}     
For our test We use 2 virtual machine equipped with Accumulo 1.7 with 80GB of virtual hard disk and 8GB of ram, the pre-processed phase took 17 minutes. 

\section{performance analysis}
To test our approach we then coded a Java class that performs DNA queries over our Accumulo datastore using Accumulo internal scans. To test the performance we did a first rough estimation by performing 10000 scan with a single process on the datastore. 

For every scan we first create a random pattern with the four letters $A,C,T,G$ of a random pattern-length between 1 and 100. Then for every scan we perform the reply time in milliseconds, as difference between the time the scan reply and the time the scan was sent to Accumulo. Last we simply evaluate outcome of a scan to be value 0 meaning no pattern matching and 1 at least one matching. In our test we do not evalute the occurrence of the pattern in the suffix array. 

The reply time in certain case was above the minimum unit length, i.e a millisecond. In this case we set the time in one milliseconds. The result was stored in a MySQL database outside our Accumulo cluster, for performance analysis. A single table was used to store all the results and a portion is the one in TABLE II

\begin{table}[!htbp] \centering 
  \caption{} 
  \label{} 
\begin{tabular}{@{\extracolsep{5pt}} cccccc} 
\\[-1.8ex]\hline 
\hline \\[-1.8ex] 
 & user & pattern & reply(mms) & outcome & pattern length \\ 
\hline \\[-1.8ex] 
1 & singlethread & CGC... & $73$ & $1$ & $28$ \\ 
2 & singlethread & CGA... & $20$ & $0$ & $34$ \\ 
3 & singlethread & CCT... & $20$ & $0$ & $50$ \\ 
4 & singlethread & GAC... & $11$ & $0$ & $78$ \\ 
\hline \\[-1.8ex] 
\end{tabular} 
\end{table} 

We also define a column name user for performing multi-thread query and group the single thread scans. Using the R statistical framework \cite{ripley2001r}, we underline in TABLE III a first estimation of a single user process 10000 scans.

\begin{table}[!htbp] \centering 
  \caption{} 
  \label{} 
\begin{tabular}{@{\extracolsep{5pt}}lccccc} 
\\[-1.8ex]\hline 
\hline \\[-1.8ex] 
Statistic & \multicolumn{1}{c}{N} & \multicolumn{1}{c}{Mean} & \multicolumn{1}{c}{St. Dev.} & \multicolumn{1}{c}{Min} & \multicolumn{1}{c}{Max} \\ 
\hline \\[-1.8ex] 
milliseconds & 10,000 & 2.790 & 3.639 & 1 & 41 \\ 
outcome & 10,000 & 0.072 & 0.259 & 0 & 1 \\ 
\hline \\[-1.8ex] 
\end{tabular} 
\end{table} 

For a single process the reply time is

And the distribution of the reply time in millisecond for one single process

\begin{figure}[htbp]
	\centering
		\includegraphics[width=0.5\textwidth]{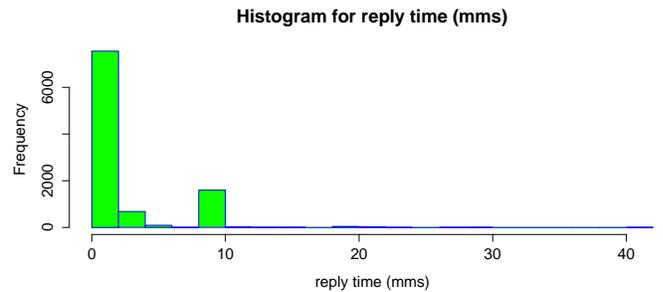}
	\caption{Single user scans reply frequency}
	\label{enrich}
\end{figure}

Next we move on a multithread approach performing the same 10000 scans with random-length, random-four-digit string,  over the Accumulo datastore, using 50 different computational threads, in order to simulate 50 users that perform 10000 scans each. The results are displayed in TABLE IV

\begin{table}[!htbp] \centering 
  \caption{} 
  \label{} 
\begin{tabular}{@{\extracolsep{5pt}}lccccc} 
\\[-1.8ex]\hline 
\hline \\[-1.8ex] 
Statistic & \multicolumn{1}{c}{N} & \multicolumn{1}{c}{Mean} & \multicolumn{1}{c}{St. Dev.} & \multicolumn{1}{c}{Min} & \multicolumn{1}{c}{Max} \\ 
\hline \\[-1.8ex] 
milliseconds & 526,395 & 5.258 & 7.667 & 1 & 771 \\ 
outcome & 526,395 & 0.080 & 0.271 & 0 & 1 \\ 
pattern length & 526,395 & 49.968 & 28.659 & 1 & 99 \\ 
\hline \\[-1.8ex] 
\end{tabular} 
\end{table}

with the following correlation in TABLE V

\begin{table}[!htbp] \centering 
  \caption{} 
  \label{} 
\begin{tabular}{@{\extracolsep{5pt}} cccc} 
\\[-1.8ex]\hline 
\hline \\[-1.8ex] 
 & milliseconds & outcome & pattern length \\ 
\hline \\[-1.8ex] 
milliseconds & $1$ & $$-$0.0002$ & $0.013$ \\ 
outcome & $$-$0.0002$ & $1$ & $$-$0.469$ \\ 
pattern length & $0.013$ & $$-$0.469$ & $1$ \\ 
\hline \\[-1.8ex] 
\end{tabular} 
\end{table} 

So as for the single process scan it seems that there is no correlation between the pattern length and the replay time neither between the pattern length. The interesting fact is that the slowest reply is of 771 mms while the mean is of 5 milliseconds. This result are in line with the fastest performances tested on and Acculumo large scale infrastructure \cite{kepner2014achieving}.

\section{Conclusion}
In this paper we explore the possibility to use an high performance big table distributed storage to store and query DNA string implementing the well-known suffix array. 

With the booming of the so called big data DNA \cite{howe2008big}, the need of fast performing online DNA query tools with exact match will increase in the next years. We think that our approach could help  understanding what will be the best way to deal with the future astonishing amount of data coming from the personalized medicine.

Apart from the pre-processing time for creating the structure that needs to be done once, we have seen that the performances of a single process query random DNA string of random length are very high. This is basically a confirmation of something already shown in previous works \cite{kepner2014achieving}.   

Obviously the work could be easily extended by extending some basic assumptions. The first one is that we have only a portion of a human DNA, not the whole. However it is some research seems to suggests that this problem can be easily bypassed using a big data structure. The problem of the storage size and the query performances could also be explored better even if some research have already been conducted
\cite{sawyer2013understanding}.

We perform single thread and multi-thread queries and statistically evaluate the performances. In this case we started with a single DNA-chromosome however it will be interesting to add to our rows also the biological organism to to allow what Blast \cite{blast_website} allow today. Blast is able not only to evaluate patter-matching in a single DNA string but also to send a DNA string and find the matching between different living being. Our infrastructure does not allow a big farm implementation of such idea but it will be interesting to see this idea in action.

Concluding we need to tell that while we was testing our system a similar research have been conducted by Dodson,S. Ricke et Alt.  \cite{dodson2014genetic} from MIT Lincoln Laboratory, but not using our approach. The authors instead started from Blast \cite{blast_website} that is a probabilistic algorithm for pattern matching \cite{tatusova1999blast} and used Accumulo and D4M Big Data
Approach \cite{kepner2012dynamic} to boost up the computational time. 

We leave to the willing readers all these suggestions that, for us, will be subject to further investigations.




%
%
%


\end{document}